\documentclass[%
 reprint,
 superscriptaddress,
 amsmath,amssymb,
 aps,longbibliography,
]{revtex4-1}

\usepackage{graphicx}
\usepackage{dcolumn}
\usepackage{bm}

\usepackage{blindtext}
\usepackage{appendix}
\usepackage{lipsum}
\usepackage{xr}
\usepackage{xcolor}
\usepackage{siunitx}

\usepackage[colorlinks]{hyperref}
\hypersetup{%
    plainpages=true,
    breaklinks=true,
    hypertexnames=false,
    pageanchor=true,
    colorlinks=true,
    linkcolor={black},
    citecolor={blue},
    urlcolor={blue},
    anchorcolor={black}
}
\begin{document}
\renewcommand{\figurename}{Fig.}

\preprint{APS/123-QED}

\title{Waveguide Quantum Electrodynamics with Giant Superconducting Artificial Atoms}%

\author{Bharath Kannan}
\email{bkannan@mit.edu}
\affiliation{Research Laboratory of Electronics, Massachusetts Institute of Technology, Cambridge, MA 02139, USA}

\affiliation{Department of Electrical Engineering and Computer Science, Massachusetts Institute of Technology, Cambridge, MA 02139, USA}

\author{Max J. Ruckriegel}
\affiliation{Research Laboratory of Electronics, Massachusetts Institute of Technology, Cambridge, MA 02139, USA}

\author{Daniel L. Campbell}
\affiliation{Research Laboratory of Electronics, Massachusetts Institute of Technology, Cambridge, MA 02139, USA}

\author{Anton Frisk Kockum}
\affiliation{Wallenberg Centre for Quantum Technology, Department of Microtechnology and Nanoscience, Chalmers University of Technology, 412 96 Gothenburg, Sweden}

\author{Jochen Braum\"uller}
\affiliation{Research Laboratory of Electronics, Massachusetts Institute of Technology, Cambridge, MA 02139, USA}

\author{David Kim}
\affiliation{MIT Lincoln Laboratory, 244 Wood Street, Lexington, MA 02420, USA}

\author{Morten Kjaergaard}
\affiliation{Research Laboratory of Electronics, Massachusetts Institute of Technology, Cambridge, MA 02139, USA}

\author{Philip Krantz}
\altaffiliation[Present address: ]{Wallenberg Centre for Quantum Technology, Department of Microtechnology and Nanoscience, Chalmers University of Technology, 412 96 Gothenburg, Sweden}
\affiliation{Research Laboratory of Electronics, Massachusetts Institute of Technology, Cambridge, MA 02139, USA}

\author{Alexander Melville}
\affiliation{MIT Lincoln Laboratory, 244 Wood Street, Lexington, MA 02420, USA}

\author{Bethany M. Niedzielski}
\affiliation{MIT Lincoln Laboratory, 244 Wood Street, Lexington, MA 02420, USA}

\author{Antti Veps\"al\"ainen}
\affiliation{Research Laboratory of Electronics, Massachusetts Institute of Technology, Cambridge, MA 02139, USA}

\author{Roni Winik}
\affiliation{Research Laboratory of Electronics, Massachusetts Institute of Technology, Cambridge, MA 02139, USA}

\author{Jonilyn Yoder}
\affiliation{MIT Lincoln Laboratory, 244 Wood Street, Lexington, MA 02420, USA}

\author{Franco Nori}
\affiliation{Theoretical Quantum Physics Laboratory, RIKEN Cluster for Pioneering Research, Wako-shi, Saitama 351-0198, Japan}
\affiliation{Department of Physics, The University of Michigan, Ann Arbor, Michigan  48109-1040, USA}

\author{Terry P. Orlando}
\affiliation{Research Laboratory of Electronics, Massachusetts Institute of Technology, Cambridge, MA 02139, USA}
\affiliation{Department of Electrical Engineering and Computer Science, Massachusetts Institute of Technology, Cambridge, MA 02139, USA}

\author{Simon Gustavsson}
\affiliation{Research Laboratory of Electronics, Massachusetts Institute of Technology, Cambridge, MA 02139, USA}

\author{William D. Oliver}
\affiliation{Research Laboratory of Electronics, Massachusetts Institute of Technology, Cambridge, MA 02139, USA}
\affiliation{Department of Electrical Engineering and Computer Science, Massachusetts Institute of Technology, Cambridge, MA 02139, USA}
\affiliation{MIT Lincoln Laboratory, 244 Wood Street, Lexington, MA 02420, USA}
\affiliation{Department of Physics, Massachusetts Institute of Technology, Cambridge, MA 02139, USA}

\maketitle
\begin{figure*}[t]
    \centering
    \includegraphics[width=\textwidth]{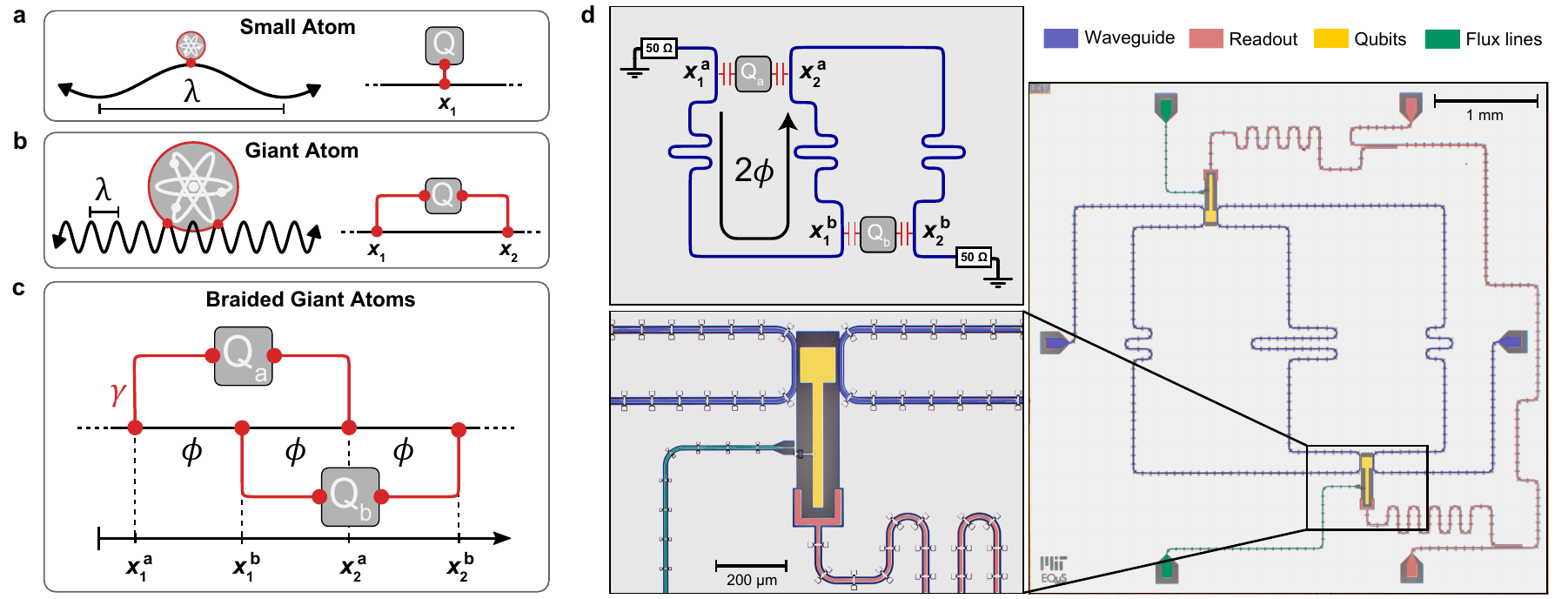}
    \caption{\textbf{Giant atoms with superconducting qubits.}\ \textbf{a)} A schematic diagram of a small atom. The small atom is treated as a point-like object since it is much smaller than the wavelength, $\lambda$, of the mode it interacts with. \textbf{b)} A giant atom that is formed by coupling a small atom to a mode at two discrete distant locations. \textbf{c)} The configuration of two braided giant atoms that are coupled to a waveguide twice with equal strength. The phase difference between coupling points $\phi = \omega(x_1^b - x_1^a)/v = \omega(x_2^a - x_1^b)/v = \omega(x_2^b - x_2^a)/v$ is varied by tuning the qubit frequencies $\omega$. \textbf{d)} A schematic diagram (top left) and false-colored optical micrograph image of a device in the configuration shown in (c). Each qubit (yellow) has a readout resonator (red) and flux line (green) for independent readout and flux control. The central waveguide (blue) is terminated to $50\,\Omega$.}
    \label{fig:figure1}
\end{figure*}

\textbf{Models of light-matter interactions typically invoke the dipole approximation \cite{D.F.WallsandG.J.Milburn2008,kockum2019quantum}, within which atoms are treated as point-like objects when compared to the wavelength of the electromagnetic modes that they interact with. However, when the ratio between the size of the atom and the mode wavelength is increased, the dipole approximation no longer holds and the atom is referred to as a ``giant atom'' \cite{FriskKockum2014,kockum2019quantum}. Thus far, experimental studies with solid-state devices in the giant-atom regime have been limited to superconducting qubits that couple to short-wavelength surface acoustic waves \cite{Manenti2017,Bolgar2018,Moores2018,Sletten2019,MartinV.GustafssonThomasArefAntonFriskKockumMariaK.EkstromGoranJohansson2014,Andersson2019,Andersson2019_2}, only probing the properties of the atom at a single frequency. Here we employ an alternative architecture that realizes a giant atom by coupling small atoms to a waveguide at multiple, but well separated, discrete locations. Our realization of giant atoms enables tunable atom-waveguide couplings with large on-off ratios and a coupling spectrum that can be engineered by device design \cite{FriskKockum2014}. We also demonstrate decoherence-free interactions between multiple giant atoms that are mediated by the quasi-continuous spectrum of modes in the waveguide-- an effect that is not possible to achieve with small atoms \cite{Kockum2018}. These features allow qubits in this architecture to switch between protected and emissive configurations in situ while retaining qubit-qubit interactions, opening new possibilities for high-fidelity quantum simulations and non-classical itinerant photon generation \cite{Garcia-Alvarez2015,Gonzalez-Tudela2015}}.

 Devices where atoms are directly coupled to waveguides are well described by waveguide quantum electrodynamics (wQED). Superconducting circuits offer an ideal platform to implement and explore the physics of wQED due to the achievable strong coupling between atomic and photonic degrees of freedom. In superconducting wQED, artificial atoms are coupled to the continuum of propagating electromagnetic modes of a one-dimensional (1D) microwave transmission line \cite{Roy2017,Gu2017}. The size of these atoms is typically much smaller than the wavelength of the modes in the transmission line, as depicted in Fig.~\ref{fig:figure1}a. Superconducting artificial atoms can be engineered to spontaneously emit most of their excitations as propagating photons in the waveguide, as opposed to other emission channels in the system. This has enabled a diverse and rich set of phenomena to be experimentally observed, such as resonance fluorescence \cite{Astafiev2010,Hoi2011, Hoi2013,Hoi2015}, collective Lamb shifts \cite{Wen2019}, and Dicke super- and subradiance \cite{Dicke1954,VanLoo2013,Mirhosseini2019}. Very recently, in a work that is complementary to ours, electromagnetically induced transparency was reported using a giant superconducting artificial atom \cite{vadiraj2020engineering}. Additionally, the 1D nature of the modes enables the waveguide to mediate long-range interactions between atoms \cite{Lalumiere2013,VanLoo2013}. 
  
  A central limitation of wQED is the ever-present and strong dissipation of qubits into the waveguide. As a result, it is difficult to prepare many-body states of the qubits. In the context of wQED, the ability to initialize and entangle qubits with high fidelity can aid in a variety of applications. For instance, using the tunable qubit-waveguide coupling offered by giant atoms, one can perform digital-analog quantum simulations of fermion-fermion scattering via a bosonic bath \cite{Lamata2018,Garcia-Alvarez2015}. Additionally, the ability to generate non-classical itinerant microwave photons from qubit emitters that are directly coupled to a waveguide \cite{Gonzalez-Tudela2015,pol2017} can be applied in quantum communication and teleportation protocols \cite{Kimble2008,OBrien2009}. Recently, correlated dissipation between qubits was used to prepare entangled dark states that are decoupled from the waveguide \cite{Mirhosseini2019}. However, only a fraction of the total qubit Hilbert space is protected from dissipation with this approach. 

In this work, we overcome this obstacle by using an appropriate arrangement of giant atoms along a waveguide that protects the \textit{entire} Hilbert space of the qubits while preserving waveguide-mediated qubit-qubit interactions \cite{Kockum2018}. We show how the geometry of the giant atoms can be altered to engineer the qubit-waveguide and qubit-qubit couplings. We then use this engineered spectrum to prepare entangled states of the qubits and show that, even in the presence of a waveguide, giant atoms can be used to perform coherent quantum operations. 

Our devices consist of two frequency-tunable transmon qubits \cite{Koch2007}, $Q_a$ and $Q_b$, each of which are coupled to a superconducting resonator for dispersive readout \cite{Blais2004}. The signal from the resonators is first amplified by a travelling wave parametric amplifier \cite{Macklin307} to maximize the readout efficiency. We construct giant atoms of the form shown in Fig.~\ref{fig:figure1}b by coupling the qubits to a $50\,\Omega$ coplanar waveguide at two locations. The qubits are coupled to the waveguide in a braided manner, as shown in Figs.~\ref{fig:figure1}c and~\ref{fig:figure1}d, such that the first coupling point of qubit $Q_b$ at $x^b_1$ is centered between the coupling points of qubit $Q_a$ at $x^a_1$ and $x^a_2$. Photons in the waveguide will experience a phase shift $\phi=2\pi\Delta x/\lambda(\omega)$ between neighboring coupling points, depending on the spatial separation between successive coupling points $\Delta x$ and the wavelength $\lambda(\omega)$ for modes at the qubit frequencies $\omega = \omega_{a,b}$. The phase $\phi$ can then be tuned in situ by changing the qubit frequencies, and thereby the wavelength $\lambda(\omega) = 2\pi v/\omega$, where $v$ is the speed of light in the waveguide. The device is designed symmetrically such that $\Delta x=x^b_1-x^a_1=x^a_2-x^b_1=x^b_2-x^a_2$. Moreover, the individual physical couplings between the qubits and the waveguide are designed to be of equal strength $\gamma(\omega) \approx \gamma_0 (\omega/\omega_0)^2$, where $\gamma_0$ is the coupling strength at a reference frequency $\omega_0$. Under these conditions, the interaction Hamiltonian between the qubits and the waveguide is

\begin{align}
    H_I &= \hbar\sum_{k,j} J(\omega_k)\ \sigma^{(j)}_x [a^{\dag}_{k,L}A_j(\omega_k)+a^{\dag}_{k,R}A_j^*(\omega_k)+\text{h.c.}],
\end{align}
where the sum over $k$ indexes the modes in the waveguide and $j \in \{a, b\}$ is summed over the qubits. \textcolor{black}{The coupling strength $J(\omega_k)$ between the qubits and the $k^{\textrm{th}}$ mode of the waveguide determines the physical qubit-waveguide coupling at the qubit frequency $\omega$}
\begin{equation}
    \textcolor{black}{\gamma(\omega) = 4\pi J(\omega)^2 D(\omega),}
\end{equation}
\textcolor{black}{where $D(\omega)$ is the density of states in the waveguide \cite{FriskKockum2014}.} We further distinguish between left- and right-propagating waveguide modes with the subscripts $L, R$ in the field creation operators $a^{\dag}_{L,k}$ and $a^{\dag}_{R,k})$. The qubit Pauli $X$ operators $\sigma^{(j)}_x$ mediate the excitation exchange to and from the qubits. The effect of coupling the qubits at multiple points along the waveguide is accounted for by modifying the complex coupling amplitude
\begin{equation}
    A_j(\omega) = \sum_n e^{-i\omega x^{(j)}_n/v}.
\end{equation}
This is the sum over all phase factors for each connection point $n \in \{1,2\}$ of qubit $Q_j$ at positions $x^{(j)}_n$ along the waveguide. This frequency-dependent amplitude captures all interference effects between the coupling points of both giant atoms.

\begin{figure}[t]
    \centering
    \includegraphics[width=3.3in]{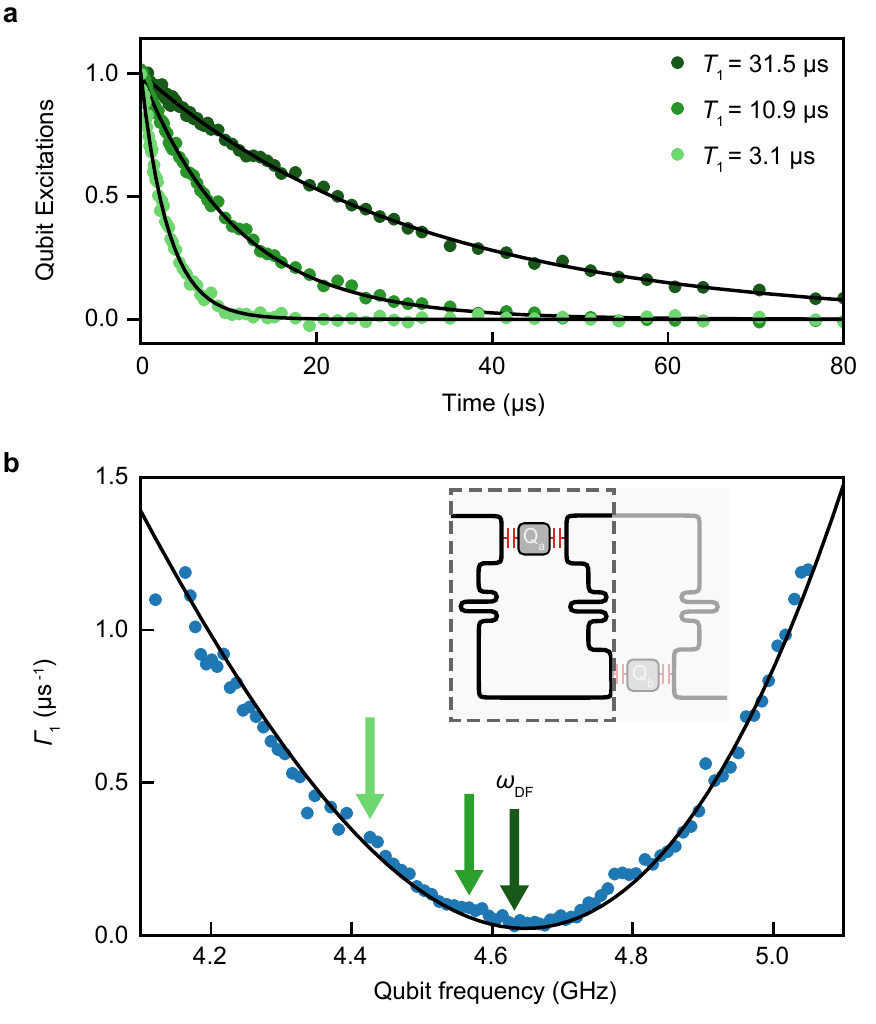}
    
    \caption{\textbf{Tunable coupling for a single giant atom.}\ \textbf{a)} $T_1$ traces and fits for qubit $Q_a$ at three different qubit frequencies. The fitted $T_1$ values are given in the top right.  \textbf{b)} The relaxation rate $\mathit{\Gamma}_1 = T_1^{-1}$ for qubit $Q_a$  as a function of its frequency with the fit to theory (solid line). The three green arrows correspond to the relaxation rate for the $T_1$ traces shown in (a). The qubit decouples from the waveguide at a decoherence-free frequency $\omega_{\textrm{DF}}/2\pi = \SI{4.645}{GHz}$ with a lifetime of $T_1 = \SI{31.5}{\micro s}$.}
    \label{fig:fig2}
\end{figure}

\begin{figure}[t]
    \centering
    \includegraphics[width=3.3in]{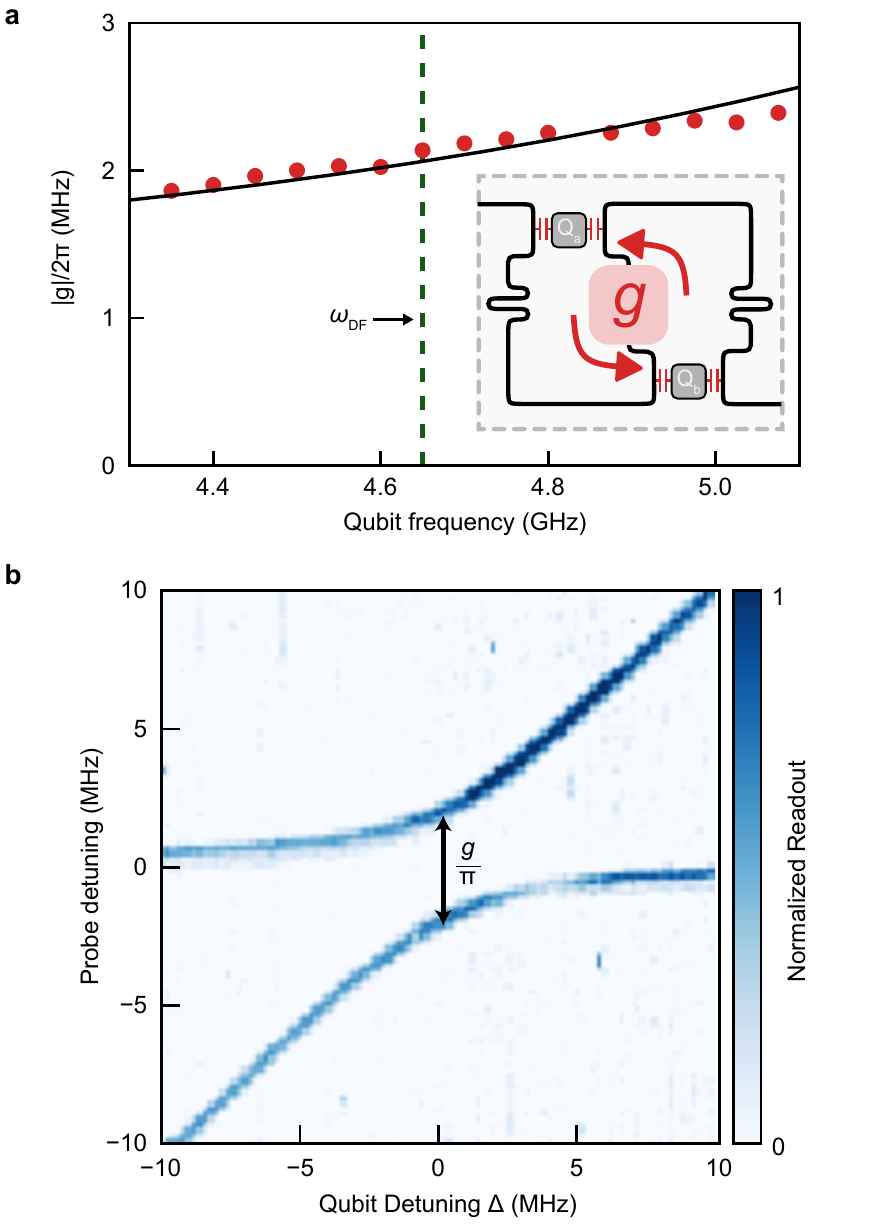}
    
    \caption{\textbf{Decoherence-free interactions between two giant atoms.}\ \textbf{a)} The measured exchange-interaction strengths between the qubits from avoided level crossings as a function of the qubit frequencies. The fit to theory (solid line) is plotted with the data (red dots) and the decoherence-free frequency is shown (dashed line). \textbf{b)} The qubit spectroscopy of an avoided level crossing centered at the decoherence-free frequency $\omega_{\textrm{DF}}/2\pi = \SI{4.645}{GHz}$. The frequency of the applied drive is swept to probe the qubit frequencies for multiple qubit detunings $\Delta$. The readout signals for each qubit are normalized and summed together to obtain both branches of the crossing. Even though the qubits are decoupled from spontaneous emission into the waveguide, a qubit exchange interaction that is mediated by virtual photons in the waveguide is measured to be $g/2\pi = \SI{2}{MHz}$.}
    \label{fig:fig3}
\end{figure}

To characterize this system, we first detune the qubits away from each other such that $|\omega_a - \omega_b| \gg \gamma$ and probe the properties of the giant atoms independently. The relaxation rates $\mathit{\Gamma}_1 = T_1^{-1}$ for the qubits are obtained by measuring the qubit energy relaxation time $T_1$ as a function of the qubit frequency. We show three representative $T_1$ traces in Fig.~\ref{fig:fig2}a for different qubit frequencies. The relaxation rates over a wide range of qubit frequencies are shown in Fig.~\ref{fig:fig2}b. For each qubit, the expression for $\mathit{\Gamma}_1$ is \cite{Kockum2018}
\begin{equation}
\label{eq:2ptgamma}
    \mathit{\Gamma}_1 = 2\gamma(\omega)[1+\mathrm{cos}(2\phi)] + \gamma_{\textrm{nr}},
\end{equation}
where $\gamma_{nr}$ is the intrinsic ``non-radiative" qubit lifetime limited by decay channels other than loss into the waveguide \cite{lu2019}. The qubit relaxation rate depends strongly on the phase $\phi$ and, thereby, the qubit frequency. This is due to quantum interference between photons emitted into the waveguide at the two coupling points of the qubit. At $2\phi = \pi$, the qubit will effectively decouple from relaxation into the waveguide due to destructive interference. We observe this phenomenon when the qubit is set to the decoherence-free frequency $\omega_{\textrm{DF}}/2\pi = \SI{4.645}{GHz}$. Here, the lifetime of the qubit is measured to be $T_1 = \SI{31.5}{\micro s}$, consistent with the typical lifetime of tunable transmon qubits in a cavity QED architecture \cite{Kjaergaard2019} despite its strong physical couplings to the continuum of modes in a waveguide. This indicates that the qubit relaxation into the waveguide is largely suppressed. The fitted theory curve Eq.~(\ref{eq:2ptgamma}) in Fig.~\ref{fig:fig2}b is in good agreement with the measured data, from which $\gamma_0/2\pi = \SI{2}{MHz}$ and $\gamma_{nr}/2\pi \approx \SI{0.03}{MHz}$ at $\omega_0 = \omega_\textrm{DF}$ are inferred. \textcolor{black}{We perform the same measurement and fit on $Q_b$ to verify that its $\gamma_0$ and $\omega_{DF}$ are identical to that of $Q_a$.}

\begin{figure*}[t]
    \centering
    \includegraphics[width=\textwidth]{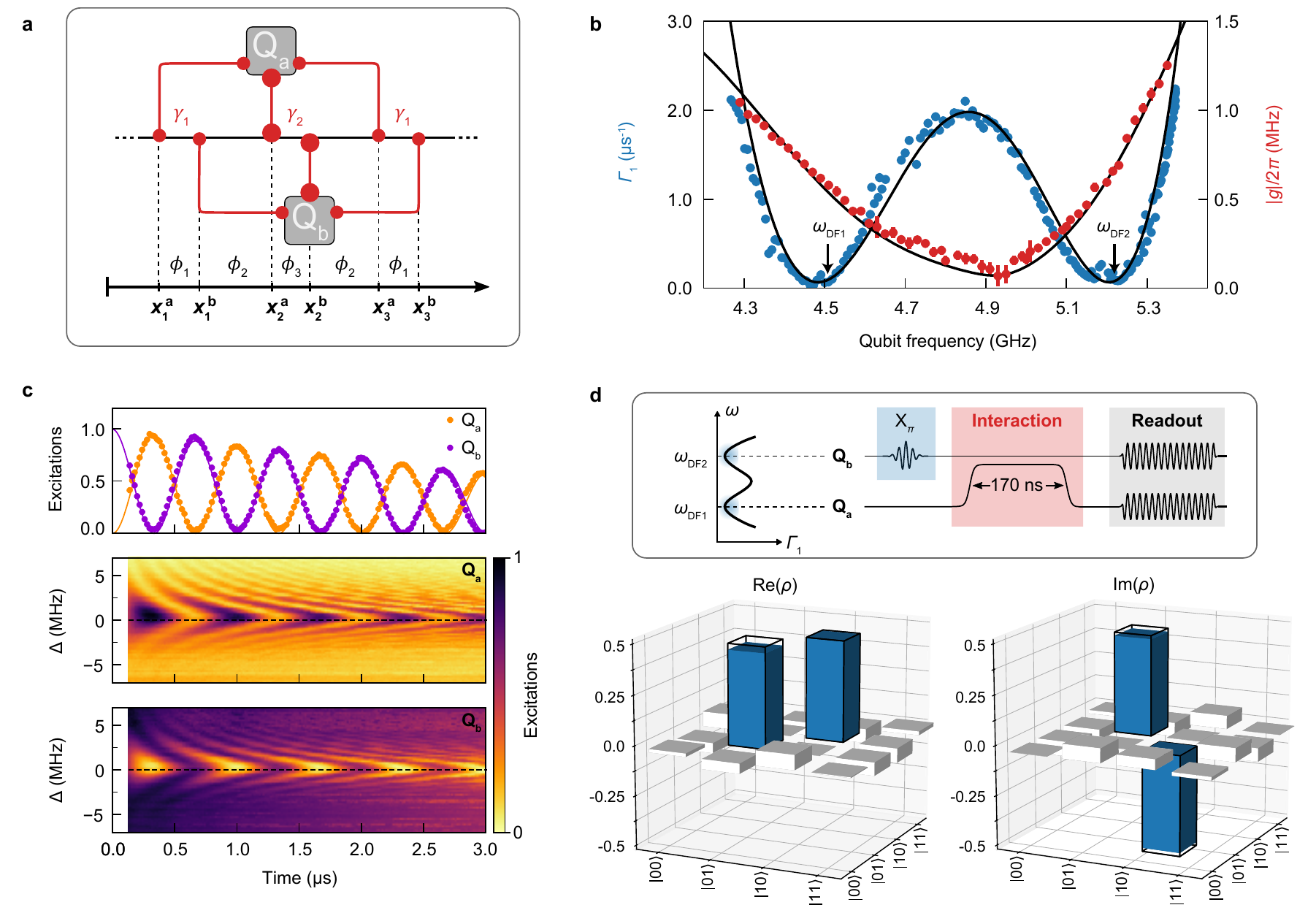}
    \caption{\textbf{Entangling qubits in wQED with engineered giant-atom geometries.} \textbf{a)} A schematic diagram of a giant-atom device with two qubits coupled to the waveguide at three points. The first and last coupling point for each qubit are designed to be equal in strength ($\gamma_1$) and the central coupling point is larger ($\gamma_2 > \gamma_1$). The ratios $\phi_1/\phi_2$, $\phi_1/\phi_3$, and $\phi_2/\phi_3$ are fixed in hardware by the relative length of the waveguide segments. \textbf{b)} The measured and fitted \textcolor{black}{single-qubit} relaxation rates (blue) and \textcolor{black}{resonant $\Delta=0$ exchange interaction strengths between $Q_a$ and $Q_b$ }(red) with fitting error bars for the configuration of giant atoms shown in (a). The design has two decoherence-free frequencies at $\omega_{\textrm{DF1}}/2\pi = \SI{4.51}{GHz}$ and $\omega_{\textrm{DF2}}/2\pi = \SI{5.23}{GHz}$. The coupling strengths inferred from the fits are $\gamma_1/2\pi = \SI{1.58}{MHz}$ and $\gamma_2/2\pi = \SI{3.68}{MHz}$ at a reference frequency $\omega_0/2\pi = \SI{5.23}{GHz}$. \textbf{c)} The time-domain chevron pattern demonstrating the exchange coupling between the qubits at the upper decoherence-free frequency $\omega_{\textrm{DF2}}$ as a function of interaction time and qubit-qubit detuning $\Delta$. \textcolor{black}{The excitation will swap maximally with a period $\pi/g = \SI{680}{ns}$ when $\Delta = 0$ and is suppressed as $|\Delta|$ increases}. \textbf{d)} The pulse sequence for preparing the entangled state $(|01\rangle - i |10\rangle)/\sqrt{2}$. $Q_a$ is placed at $\omega_{\textrm{DF1}}$ while $Q_b$ is placed at $\omega_{\textrm{DF2}}$ and excited with a $\pi$-pulse. $Q_a$ is then brought onto resonance $\Delta = 0$ and interacts with $Q_b$ for a time $\pi/4g = \SI{170}{ns}$. $Q_a$ is then returned to $\omega_{\textrm{DF1}}$ and tomography readout pulses are applied. The real and imaginary parts of the qubit density matrix obtained for this pulse sequence are shown below with matrix elements that are ideally non-zero shaded in blue. The state preparation fidelity is $94\%$. }
    \label{fig:fig4}
\end{figure*}

Next, we investigate waveguide-mediated qubit-qubit interactions. We extract the interaction strength from a fit to the avoided level crossings observed in qubit spectroscopy, performed for various qubit frequencies. The measured and fitted interaction strengths are shown in Fig.~\ref{fig:fig3}a. Of particular note is the finite exchange interaction at the decoherence-free frequency $\omega_{\textrm{DF}} = \omega_a = \omega_b$, for which the avoided level crossing is shown in Fig.~\ref{fig:fig3}b. The waveguide continues to mediate interactions between the qubits even though the qubits are decoupled from relaxation into it. This can be understood by considering the device geometry. With small atoms, atomic exchange interactions are mediated by virtual photons in the waveguide that are emitted and absorbed by different atoms. The strength of these interactions $g = \gamma \sin(\theta)/2$ is periodic in the phase delay $\theta$ between the coupling points of the small atoms \cite{Lalumiere2013}. A similar effect occurs with multiple giant atoms coupled to a waveguide. In our device, the total exchange interaction is obtained by summing the contribution from all four possible pairs of coupling points between the two qubits. For our braided configuration, the virtual photons emitted and absorbed by the three consecutive coupling pairs ($x^a_1 \leftrightarrow x^b_1$), ($x^b_1 \leftrightarrow x^a_2$), and ($x^a_2 \leftrightarrow x^b_2$) will contribute $\gamma\sin(\phi)/2$ to the interaction strength. The pair formed by the two outermost points ($x^a_1 \leftrightarrow x^b_2$) will contribute $\gamma\sin(3\phi)/2$. The interaction strength between the braided giant atoms in our device is then given by 
\begin{equation}
    g = \frac{\gamma(\omega)}{2} \left[ 3 \sin\left(\phi\right) + \sin\left(3\phi\right) \right].
    \label{eq:g}
\end{equation}
 At the decoherence-free frequency, we have $\phi = \pi/2$ and, therefore, a net interaction strength $g = \gamma$. This effect, where $\mathit{\Gamma}_1=0$ and $g\neq0$, is unique to braided configurations of giant atoms \cite{Kockum2018} and is a necessary condition for high-fidelity state preparation of qubits coupled to a waveguide.

The spectra presented in Figs.~\ref{fig:fig2}b and~\ref{fig:fig3}a are largely determined by the geometry of the device. That is, one can vary the number of coupling points, their relative coupling strengths, and distance between them to tailor the frequency dependence of the qubit-waveguide and qubit-qubit coupling rates. As a demonstration of this, we probe a device with a new geometry such that qubit relaxation and exchange can be simultaneously suppressed. We then use this property to perform entangling gates between the qubits. The device consists of two qubits that are each coupled to a waveguide at three locations, and is parameterized by three distinct phase delays ($\phi_1$, $\phi_2$, and $\phi_3$), as shown in Fig.~\ref{fig:fig4}a. The device is designed such that the physical qubit-waveguide coupling strengths are $\gamma_1$ at $x^{a/b}_1$ and $x^{a/b}_3$ and $\gamma_2$ at $x^{a/b}_2$. The relaxation times and qubit-qubit coupling strengths are \textcolor{black}{(see Methods for derivation)}
\begin{equation}
\begin{aligned}
\label{eq:giantAtoms:3pointgamma}
    \mathit{\Gamma}_1&=\gamma_2 + 2\gamma_1\left[1 + \cos{(\phi_1+2\phi_2+\phi_3)} \right] \\
    &+2\sqrt{\gamma_1 \gamma_2} \left[\cos{(\phi_1+\phi_2)} + \cos{(\phi_2+\phi_3)}\right] + \gamma_{\textrm{nr}},
\end{aligned}
\end{equation}
\begin{equation}
\begin{aligned}
    \label{eq:giantAtoms:3pointg}
    g&=\sqrt{\gamma_1\gamma_2}\left[\sin(\phi_2)+\sin(\phi_1+\phi_2+\phi_3)\right] \\
    &+\frac{\gamma_1}{2}\left[2\sin(\phi_1)+\sin(2\phi_2+\phi_3)+\sin(2\phi_1+2\phi_2+\phi_3)\right] \\
   &+\frac{\gamma_2}{2}\sin(\phi_3).
\end{aligned}
\end{equation}
\textcolor{black}{From length measurements and microwave simulations of the different sections of the waveguide, we find the ratios between the phase delays to be $\phi_1 = \phi_3 = 0.505\phi_2$. Fixing these ratios, we fit the spectra and infer} the coupling strengths to be $\gamma_1/2\pi = \SI{1.58}{MHz}$ and $\gamma_2/2\pi = \SI{3.68}{MHz}$ at a reference frequency $\omega_0/2\pi = \SI{5.23}{GHz}$. With this geometry, we are able to engineer the relaxation spectra for both qubits to have two decoherence-free frequencies, $\omega_{\textrm{DF1}}/2\pi = \SI{4.51}{GHz}$ and $\omega_{\textrm{DF2}}/2\pi = \SI{5.23}{GHz}$, that are separated by $\SI{720}{MHz}$. Thus, each qubit can be placed at a unique decoherence-free frequency in order to simultaneously protect them from relaxation into the waveguide and suppress exchange interactions between them. The asymmetry in the measured spectra in Fig.~\ref{fig:fig4}b is due to the frequency dependence of the individual coupling points $\gamma_{1,2} \propto \omega^2$ as well as a small deviation from the desired design ($\phi_1=0.5\phi_2$). We observe a small deviation from theory in the spectrum for $g$. Ideally, the exchange interaction would be zero at $\omega/2\pi = \SI{4.95}{GHz}$. However, from our fit, we extract a small non-zero parasitic interaction that is given by the minimum value of $|g(\omega)/2\pi| \approx \SI{70}{kHz}$. Taking this parasitic coupling into account, we find very good agreement between the overall spectra and theory.

We use this engineered spectrum with two decoherence-free frequencies in order to demonstrate that giant atoms can be used to prepare entangled states in waveguide-QED devices. When placed on resonance at $\omega_{\textrm{DF2}}$, the qubits exchange excitations at a rate $g/2\pi = \SI{735}{kHz}$. We confirm this by observing the chevron pattern formed by this excitation swap as a function of the interaction time and qubit-qubit detuning $\Delta$ in Fig.~\ref{fig:fig4}c. Using the pulse sequence shown in Fig.~\ref{fig:fig4}d, we perform an entangling $\sqrt{i\mathrm{SWAP}}$ operation and to prepare the state $(|01\rangle -i|10\rangle)/\sqrt{2}$. The density matrix $\rho$ is obtained from maximum-likelihood estimation on the two-qubit state tomography and is shown in Fig.~\ref{fig:fig4}d. After correcting for readout errors \cite{Chow2010}, we find a state preparation fidelity of $\mathrm{Tr}(\sqrt{\sqrt{\sigma}\rho\sqrt{\sigma}})^2 = 94\%$, where $\sigma$ is the ideal density matrix. This demonstrated ability to entangle qubits using purely coherent interactions is a hallmark feature of giant atoms that are coupled to a waveguide. Furthermore, since the entire two-qubit Hilbert space is protected from dissipation into the waveguide, a combination of $i\mathrm{SWAP}$ and single-qubit gates can be used to prepare any state. 

\textcolor{black}{Looking forward, giant atoms offer a new way of coupling distant qubits. Conventionally, this is accomplished by coupling qubits via a resonator. However, the lower free spectral range of longer resonators imposes a limit on the maximum detuning between qubits and increases their Purcell decay. Giant atoms in wQED are not subject to either of these constraints since the waveguide consists of a continuum of modes and the suppression of relaxation is due to interference. Therefore, these devices can be scaled to a greater number of qubits that are farther separated.} For example, the architecture studied in this work can be naturally extended to perform quantum simulations of spin-models with nearest-neighbor or all-to-all long-distance interactions \cite{Kockum2018}. These non-local interactions may also prove useful in quantum error correction schemes that go beyond the nearest-neighbor coupling that is native to superconducting qubits \cite{Campbell2017}. 

\textcolor{black}{Finally, the protection of qubits for high-fidelity control and the rapid emission of quantum information as itinerant photons are generally in competition with each other with standard resonator- or waveguide-based architectures.} We have shown that it is possible to perform high-fidelity quantum gates with qubits that are in the presence of a waveguide and to switch between highly protected and emissive configurations. This enables giant atoms to be used as a high-quality source of itinerant quantum information; qubits can be initialized and entangled while biased at decoherence-free frequencies, and subsequently release the quantum information they store into photons by tuning them to a frequency with strong qubit-waveguide coupling. Given an appropriate arrangement of qubits along the waveguide, this property can be used to generate itinerant photons with spatial entanglement \cite{Kannan} or \textcolor{black}{entangled cluster states for measurement-based quantum computing \cite{Pichler11362}.} These types of sources of non-classical photons can then be applied towards \textcolor{black}{distributing entanglement and} shuttling information in extensible quantum processors and networks.

\begin{acknowledgments}
We thank Youngkyu Sung and Amy Greene for valuable discussions.
This research was funded in part by the U.S. Department of Energy, Office of Science, Basic Energy Sciences, Materials Sciences and Engineering Division under Contract No. DE-AC02-05-CH11231 within the High-Coherence Multilayer Superconducting Structures for Large Scale Qubit Integration and Photonic Transduction program (QISLBNL);
and by the Department of Defense via MIT Lincoln Laboratory under U.S. Air Force Contract No. FA8721-05-C-0002.
B.K. gratefully acknowledges support from the National Defense Science and Engineering Graduate Fellowship program.
M.K. gratefully acknowledges support from the Carlsberg Foundation during a portion of this work. 
A.F.K. acknowledges support from the Swedish Research Council (Grant No. 2019-03696), and from the Knut and Alice Wallenberg Foundation.
F.N. acknowledges support from the Army Research Office (Grant No. W911NF-18-1-0358),
Japan Science and Technology Agency (via the Q-LEAP program, and the CREST Grant No. JPMJCR1676), the JSPS KAKENHI (Grant No. JP20H00134), the Foundational Questions Institute, and the NTT PHI Laboratory.
The views and conclusions contained herein are those of the authors and should not be interpreted as necessarily representing the official policies or endorsements of the U.S. Government. 
\end{acknowledgments}

\section*{Author Contributions}
B.K., D.L.C., A.F.K., F.N., S.G., and W.D.O. conceived and designed the experiment. B.K. and D.L.C. designed the devices. B.K. and M.J.R. conducted the measurements and B.K., M.J.R., A.F.K., J.B., and M.K. analyzed the data. D.K., A.M., B.M.N., and J.Y. performed sample fabrication. B.K. and M.J.R. wrote the manuscript. P.K., A.V., and R.W. assisted with the experimental setup. T.P.O., S.G., and W.D.O. supervised the project. All authors discussed the results and commented on the manuscript.

\bibliography{main}

\setcounter{figure}{0}
\setcounter{equation}{0}
\setcounter{table}{0}
\renewcommand\theequation{S\arabic{equation}}
\renewcommand\thefigure{S\arabic{figure}}
\renewcommand\thetable{S\arabic{table}}

\setcounter{section}{0}

\clearpage
\onecolumngrid
 \section*{Supplementary Information}
 \section{Methods}
 \subsection*{Experimental Setup}
 \begin{figure*}[h]
    \centering

    \includegraphics[width=\textwidth]{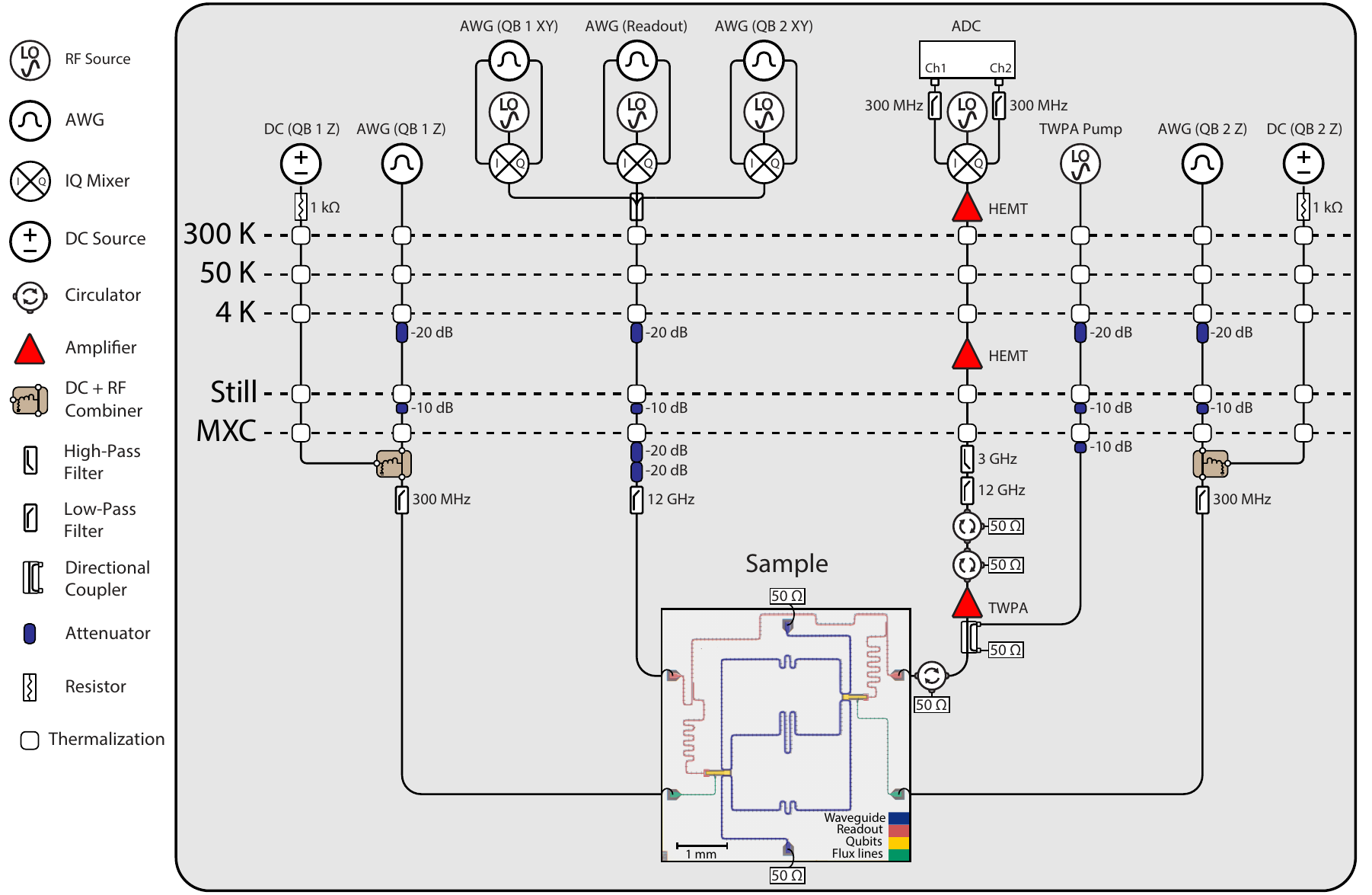}
    \caption{\textbf{Experimental Setup}. A schematic diagram of the experimental setup used to obtain the data presented in the main text. }
    \label{fig:expsetup}

\end{figure*}

The experiments are performed in a Bluefors XLD600 dilution refrigerator, capable of cooling to a base temperature of $\SI{10}{mK}$. The experimental setup used for all of the data presented in the main text is shown in Extended Data Fig.~\ref{fig:expsetup}. The readout and qubit drives are all combined at room temperature ($\SI{300}{K}$) and sent into the line that the readout resonators are coupled to. This line is attenuated by 20dB at the $\SI{4}{K}$ stage, 10dB at the still, and 40dB at the mixing chamber (MXC) to ensure proper thermalization of the line. The samples are magnetically shielded at the MXC by superconducting and Cryoperm-10 shields. To maximize the signal-to-noise ratio of the readout, a Josephson travelling wave parametric amplifier (TWPA) is used as the first amplifier in the measurement chain. The TWPA is pumped in the forward direction using a directional coupler. A circulator is placed between the sample and the TWPA to ensure that the TWPA sees a $\SI{50}{\Omega}$ impedance on both sides. The signal is then filtered with $\SI{3}{GHz}$ high-pass and $\SI{12}{GHz}$ low-pass filters. Two additional circulators are placed after the TWPA in the MXC to prevent noise from higher-temperature stages travelling back into the TWPA and the sample. High electron mobility transistor (HEMT) amplifiers are used at $\SI{4}{K}$ and room-temperature stages of the measurement chain for further amplification. The signal is then downconverted to an intermediate frequency using an IQ mixer, filtered, digitized, and demodulated. 

The central waveguide of the device to which the qubits couple is directly terminated to $\SI{50}{\Omega}$ at the device package. \textcolor{black}{This minimizes impedance mismatches that would otherwise come from external components, such as circulators and filters, which may alter the frequency dependence of the qubit-waveguide coupling strength.} The frequencies of the qubits are controlled with local flux lines. Each flux line has both DC and RF control that are combined and filtered with $\SI{300}{MHz}$ low-pass filters at the mixing chamber. The RF flux control line is attenuated by 20dB at the $\SI{4}{K}$ stage and by 10dB at the still. A $\SI{1}{k\Omega}$ resistor is placed in series with the DC voltage source to generate a DC current. 

\subsection*{Flux Cross-talk Calibration and Frequency Control}
All devices used in the experiment had local flux lines for independent flux control, and it is necessary to calibrate the cross-talk between these lines. We measure both qubit frequencies as a function of voltage applied to each flux line. The frequency spectrum of qubit $i$ from the flux line of qubit $j$ is fit to the analytical transmon frequency spectrum $f_{i,j}(V_j)$, given by \cite{Koch2007}
\begin{equation}
\label{eq:transmon}
    f_{i,j}=f_{i,\max}\sqrt[4]{d^{2}+\left(1-d^{2}\right) \cos ^{2}\left(\pi \frac{V_j}{V_{0,i,j}}-\phi_0\right)},
\end{equation}
where $d=f_{i,\min}/f_{i, \max}$ describes the asymmetry between the junctions in the transmon SQUID loop, $\phi_0$ is the offset from zero, $V_i$ is voltage applied on the flux line, and $f_{i, \min}, f_{i, \max}$ are the minimum and maximum frequencies of the transmon, respectively. The fit parameter $V_{0,i,j}$ describes the voltage required to supply a flux quantum $\Phi_0$. These values for all combinations of $i,j\in\{Q_a,Q_b\}$ can be used to construct a cross-talk matrix $S$, where the matrix elements of $S$ are given by $S_{i,j} = V_{0,i,j}^{-1}$. An example of the matrix $S$ used in the experiment is given by
\begin{equation}
    S = \left(\begin{array}{cc} 4.46^{-1} & 98.8^{-1}\\ 103.4^{-1} & 4.26^{-1} \end{array}\right) [\Phi_0/V]
\end{equation}
Here, the diagonal matrix elements correspond to the coupling between a qubit and its own local flux line, whereas the off-diagonal elements represent the coupling between a qubit and the other qubits flux line. The qubit frequencies $f_{i,i}$ from Eq.~(\ref{eq:transmon}) can also be expressed in terms of the flux applied on qubit $i$ using the relation $\Phi_i = \Phi_0V_i/V_{0,i,i}$. Thus, a desired configuration of qubit frequencies $\textbf{f}=(f_1,f_2)$ can be mapped onto a set of fluxes $\Phi=(\Phi_1,\Phi_2)$. The voltage configuration $\textbf{V}=(V_1,V_2)$ required to achieve the desired flux, and therefore qubit frequency, configuration can then be found by using the inverted cross-talk matrix $\textbf{V}=S^{-1}\Phi$.

\subsection*{Summary of Device Parameters}
\begin{figure*}[t]
    \centering

    \includegraphics[width=\textwidth]{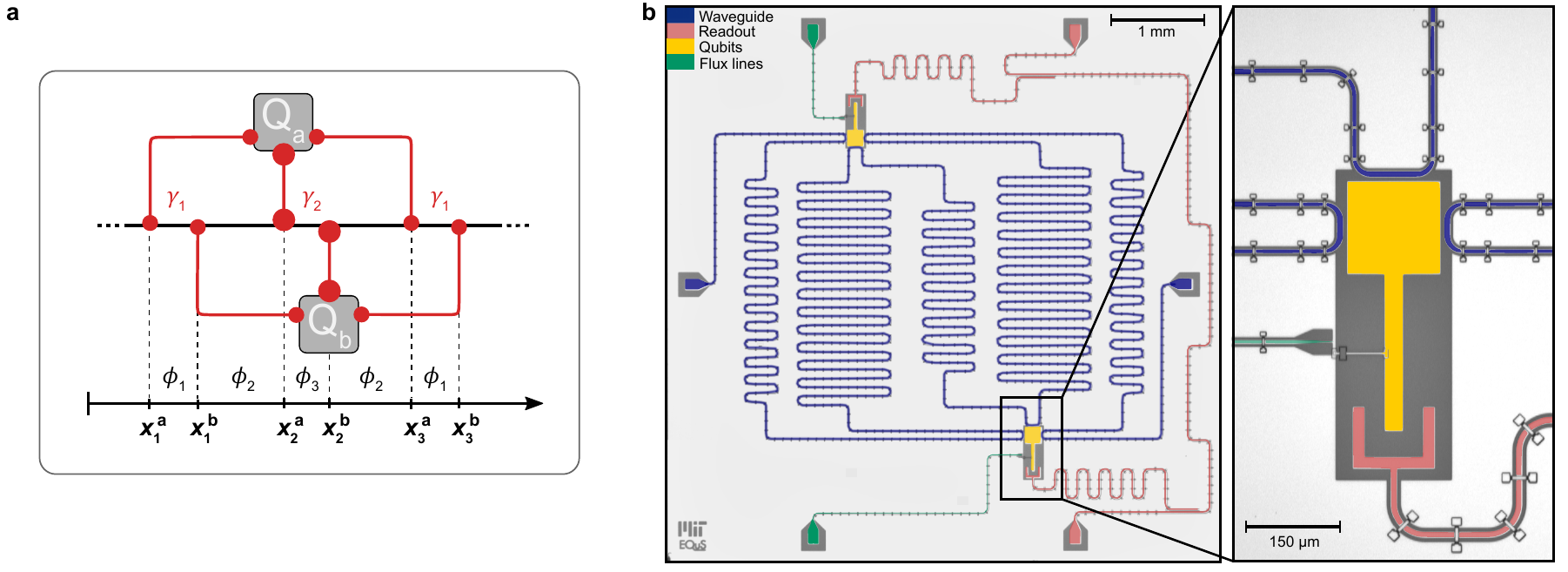}
    \caption{\textbf{Device C.} \textbf{a)} A schematic diagram of a giant-atom device with two qubits coupled to the waveguide at three points. The ratios $\phi_1/\phi_2$, $\phi_1/\phi_3$, and $\phi_2/\phi_3$ are fixed in hardware. \textbf{b)} A false-colored optical micrograph image of a device in the configuration shown in (a). Each qubit (yellow) has a readout resonator (red) and flux line (green) for independent readout and flux control. The central waveguide (blue) is terminated to $50\,\Omega$. Airbridges are placed every $\SI{80}{\micro m}$ along the waveguide to tie the ground planes together and prevent slotline modes.}
    \label{fig:3ptdevice}

\end{figure*}

\begin{table}[b]
\begin{tabular}{c|clc|cclc|lclc}
                                             & \multicolumn{3}{c|}{Device A}                                        &  & \multicolumn{3}{c|}{Device B}                                        &                      & \multicolumn{3}{c}{Device C}                                         \\ \hline
Parameter                                    & $Q_1$                 &                      & $Q_2$                 &  & $Q_1$                 &                      & $Q_2$                 &                      & $Q_1$                 &                      & $Q_2$                 \\ \hline
Max Qubit Frequency $f_{\max}$               & $\SI{4.856}{GHz}$     &                      & $\SI{4.801}{GHz}$     &  & $\SI{5.403}{GHz}$     &                      & $\SI{5.451}{GHz}$     &                      & $\SI{5.319}{GHz}$     &                      & $\SI{5.371}{GHz}$     \\
Min Qubit Frequency $f_{\min}$               & $\SI{4.126}{GHz}$     &                      & $\SI{4.109}{GHz}$     &  & $\SI{4.58}{GHz}$      &                      & $\SI{4.621}{GHz}$     &                      & $\SI{4.101}{GHz}$     &                      & $\SI{4.162}{GHz}$     \\
Junction Asymmetry $d$                       & 0.903                 &                      & 0.856                 &  & 0.848                 &                      & 0.852                 &                      & 0.771                 &                      & 0.775                 \\
Qubit Anharmonicity $\alpha/2\pi$            & $\SI{-219}{MHz}$      &                      & $\SI{-218}{MHz}$      &  & $\SI{-213}{MHz}$      &                      & $\SI{-213}{MHz}$      &                      & $\SI{-174}{MHz}$      &                      & $\SI{-175}{MHz}$      \\
\textcolor{black}{Max $T_1$}                               & $\SI{31.5}{\micro s}$ & \multicolumn{1}{c}{} & $\SI{26.1}{\micro s}$ &  & $\SI{23.5}{\micro s}$ & \multicolumn{1}{c}{} & $\SI{29.7}{\micro s}$ & \multicolumn{1}{c}{} & $\SI{19.5}{\micro s}$ & \multicolumn{1}{c}{} & $\SI{21.2}{\micro s}$ \\
$\textcolor{black}{T^*_2}$                              & $\SI{4.2}{\micro s}$  & \multicolumn{1}{c}{} & $\SI{3.6}{\micro s}$ &  & $\SI{7.9}{\micro s}$ & \multicolumn{1}{c}{} & $\SI{10.8}{\micro s}$ & \multicolumn{1}{c}{} & $\SI{5.4}{\micro s}$  & \multicolumn{1}{c}{} & $\SI{3.9}{\micro s}$  \\
Readout Resonator Frequency $f_{r}$ & $\SI{6.982}{GHz}$     &                      & $\SI{7.161}{GHz}$     &  & $\SI{6.984}{GHz}$     &                      & $\SI{7.166}{GHz}$     &                      & $\SI{7.094}{GHz}$     &                      & $\SI{7.287}{GHz}$     \\
Dispersive Shift $\chi/2\pi$                 & $\SI{0.577}{MHz}$     &                      & $\SI{0.463}{MHz}$     &  & $\SI{1.09}{MHz}$      &                      & $\SI{0.911}{MHz}$     &                      & $\SI{0.518}{MHz}$     &                      & $\SI{0.441}{MHz}$     \\
Readout Resonator Linewidth $\kappa/2\pi$    & $\SI{0.95}{MHz}$      &                      & $\SI{0.693}{MHz}$     &  & $\SI{0.971}{MHz}$     &                      & $\SI{0.698}{MHz}$     &                      & $\SI{0.575}{MHz}$     &                      & $\SI{0.491}{MHz}$    
\end{tabular}
\caption{\textbf{Device parameters.} Summary of the qubit and readout parameters for the devices used in the data presented in Figs.~\ref{fig:fig2}, \ref{fig:fig3}, and \ref{fig:fig4} of the main text. Devices A and B are of the same design and were used for the data in Figs.~\ref{fig:fig2} and \ref{fig:fig3}. Device C was used for the data in Fig.~\ref{fig:fig4}. \textcolor{black}{The $T_1$ and $T^*_2$ values are quoted at the waveguide decoherence-free frequency.} All other frequency-dependent parameters are quoted at the maximum qubit frequency.}
\label{table:devices}
\end{table}
To obtain the data presented in Figs.~\ref{fig:fig2} and \ref{fig:fig3} of the main text, we used two devices, device A and device B, that were nominally identical to each other apart from their maximum and minimum qubit frequencies. This let us obtain the coupling spectra from a range of $\SI{4.2}{GHz}$ to $\SI{5.1}{GHz}$. Device A and B were used for data points in the range of qubit frequencies between $\SI{4.2}{GHz}-\SI{4.7}{GHz}$ and $\SI{4.7}{GHz}-\SI{5.1}{GHz}$, respectively. The general qubit and readout parameters for these devices are summarized in Extended Data Table~\ref{table:devices}. Similarly, a summary of the qubit and readout parameters for the device used for the data presented in Fig.~\ref{fig:fig4} of the main text, device C, is also given in Extended Data Table~\ref{table:devices}. An optical image of device C is shown in Extended Data Fig.~\ref{fig:3ptdevice}b. Here, it can be seen that the relative coupling strength between $\gamma_1$ and $\gamma_2$ is determined by the difference in length, and therefore capacitance, between the sections of the waveguide that couple to the qubit. 

\subsection*{Entangling Pulse Sequence}

\textcolor{black}{As shown in Fig.~\ref{fig:fig4}d, we initialize the frequencies of $Q_a$ and $Q_b$ to $\omega_{\textrm{DF1}}$ and $\omega_{\textrm{DF2}}$, respectively. This protects both qubits from dissipation into the waveguide while also suppressing the exchange interaction between them with a large detuning. Next, we excite $Q_b$ with a $\pi$-pulse and bring $Q_a$ onto resonance with it for a time $\pi/4g=\SI{170}{ns}$ to entangle the qubits. We then bring $Q_a$ back to $\omega_{\textrm{DF1}}$ to turn the interaction off and perform two-qubit state tomography.}

 \section{Relaxation and Coupling Spectra for Generalized Giant Atoms}
 
 The properties of a generalized giant-atom system composed of $N$ qubits, and where the $j^{\textrm{th}}$ atom has $M_j$ coupling points, can be calculated using the SLH formalism for cascaded quantum systems \cite{Kockum2018}. The qubit-waveguide coupling strength for the $n^{\textrm{th}}$ coupling point of atom $j$ is given by $\gamma_{j_n}$ and the phase delay between the $n^{\textrm{th}}$ and $m^{\textrm{th}}$ coupling points of atoms $j$ and $k$ is given by $\phi_{j_n,k_m}$. The master equation for this setup is given by
\begin{equation}
\begin{aligned} \dot{\rho}=&-i\left[\sum_{j=1}^{N} \omega_{j}^{\prime} \frac{\sigma_{z}^{(j)}}{2}+\sum_{j=1}^{N-1} \sum_{k=j+1}^{N} g_{j, k}\left(\sigma_{-}^{(j)} \sigma_{+}^{(k)}+\sigma_{+}^{(j)} \sigma_{-}^{(k)}\right), \rho\right] \\ &+\sum_{j=1}^{N} \mathit{\Gamma}_{j} \mathcal{D}\left[\sigma_{-}^{(j)}\right] \rho \\ &+\sum_{j=1}^{N-1} \sum_{k=j+1}^{N} \mathit{\Gamma}_{\mathrm{coll}, j, k}\left[\left(\sigma_{-}^{(j)} \rho \sigma_{+}^{(k)}-\frac{1}{2}\left\{\sigma_{+}^{(j)} \sigma_{-}^{(k)}, \rho\right\}\right)+\mathrm{H.c.}\right], \end{aligned}
\end{equation}
\noindent
where $\mathcal{D}\left[O\right] \rho = O\rho O^\dag - \frac{1}{2} \{O^\dag O, \rho\}$ is the standard Lindblad dissipator and $\omega_{j}^{\prime}$ is the frequency of the $j^{\textrm{th}}$ atom with the Lamb shifts included. This master equation assumes weak coupling, $\gamma_{j_n}\ll \omega_j^{\prime}$, and a negligible travel time between coupling points \textcolor{black}{such that $\gamma_{j_n}^{-1} \gg t$, where $t$ is the travel time for photons in the waveguide between the coupling points. Note that the latter assumption places a limit on the spatial separation $\Delta x$ between coupling points, but this limit is easily satisfied in superconducting coplanar waveguides for typical values of $\gamma_{j_n}$. For example, with typical values of $\gamma_{j_n}$, the distance between the qubits would need to be on the order of $\Delta x\approx \SI{10}{m}$ to reach $\gamma_{j_n}^{-1} = t$.}

The relaxation rate $\mathit{\Gamma}_j$ of the $j^{\textrm{th}}$ atom, the exchange interaction $g_{j, k}$ between the $j^{\textrm{th}}$ and $k^{\textrm{th}}$ atoms, and the collective decay rate $\mathit{\Gamma}_{\mathrm{coll}, j, k}$ (not studied in this experiment) for the $j^{\textrm{th}}$ and $k^{\textrm{th}}$ atoms are given by
\begin{equation}
    \label{eq:gen_gamma}
\mathit{\Gamma}_{j}=\sum_{n=1}^{M_{j}} \sum_{m=1}^{M_{j}} \sqrt{\gamma_{j_{n}} \gamma_{j_{m}}} \cos \phi_{j_{n}, j_{m}},
\end{equation}
\begin{equation}
    \label{eq:gen_g}
    g_{j, k}=\sum_{n=1}^{M_{j}} \sum_{m=1}^{M_{k}} \frac{\sqrt{\gamma_{j_{n}}\gamma_{k_{m}}}}{2} \sin \phi_{j_{n}, k_{m}},
\end{equation}
\begin{equation}
    \label{eq:gen_gammacol}
\mathit{\Gamma}_{\mathrm{coll}, j, k}=\sum_{n=1}^{M_{j}} \sum_{m=1}^{M_{k}} \sqrt{\gamma_{j_{n}} \gamma_{k_{m}}} {\cos \phi_{j_{n}, k_{m}}}.
\end{equation}
Using this result, we may verify the equations used for the fits in the main text. For the first device shown in Fig.~\ref{fig:figure1} of the main text, we have $N=2$, $M_1 = M_2 = 2$, $\gamma_{1_1}=\gamma_{1_2}=\gamma_{2_1}=\gamma_{2_2}=\gamma$, and $\phi_{1_1,2_1}=\phi_{2_1,1_2}=\phi_{1_2,2_2}=\phi$. Plugging these into equations (8), (9), and (10), and adding the contribution from decay into modes other than the waveguide $\gamma_{\textrm{nr}}$, we recover the equations \ref{eq:2ptgamma} and \ref{eq:g} of the main text

\begin{equation}
    \mathit{\Gamma}_1 = 2\gamma\left[1+\mathrm{cos}(2\phi)\right] + \gamma_{\textrm{nr}},
\end{equation}
\begin{equation}
    g = \frac{\gamma}{2} \left[ 3 \sin \left(\phi\right) + \sin\left(3\phi\right) \right].
\end{equation}
Similarly, for the device presented in Fig.~\ref{fig:fig4} of the main text we have $N=2$, $M_1 = M_2 = 3$, $\gamma_{1_1}=\gamma_{1_3}=\gamma_{2_1}=\gamma_{2_3}=\gamma_1$, $\gamma_{1_2}=\gamma_{2_2} = \gamma_2$, $\phi_{1_1,2_1}=\phi_{1_3,2_3}=\phi_1$, $\phi_{1_2,2_2}=\phi_3$, and $\phi_{2_1,1_2}=\phi_{2_2,1_3}=\phi_2$. Using these values, we obtain
\begin{equation}
\begin{aligned}
\label{eq:giantAtoms:3pointgammamethods}
    \mathit{\Gamma}_1=\gamma_2 + 2\gamma_1\left[1 + \cos{(\phi_1+2\phi_2+\phi_3)} \right]+2\sqrt{\gamma_1 \gamma_2} \left[\cos{(\phi_1+\phi_2)} + \cos{(\phi_2+\phi_3)}\right] + \gamma_{\textrm{nr}},
\end{aligned}
\end{equation}
\begin{equation}
\begin{aligned}
    \label{eq:giantAtoms:3pointgmethods}
    g=\sqrt{\gamma_1\gamma_2}\left[\sin(\phi_2)+\sin(\phi_1+\phi_2+\phi_3)\right]+\frac{\gamma_1}{2}\left[2\sin(\phi_1)+\sin(2\phi_2+\phi_3)+\sin(2\phi_1+2\phi_2+\phi_3)\right]+\frac{\gamma_2}{2}\sin(\phi_3)
\end{aligned}
\end{equation}
and recover equations (\ref{eq:giantAtoms:3pointgamma}) and (\ref{eq:giantAtoms:3pointg}) of the main text.

\end{document}